\documentclass[USenglish,twocolumn]{article}

\usepackage[utf8]{inputenc}
\usepackage[big]{dgruyter}
\usepackage{subfigure}
\usepackage{array}
\usepackage{amsmath}
\usepackage{mathtools}
\usepackage{calc}
\usepackage{caption}
\usepackage{color}
\usepackage[dvipsnames]{xcolor}

\begin{document}
  
  \articletype{Research Article{\hfill}Open Access}

\author*[1]{Matti Dorsch}
\author[2]{Marilyn Latour}
\author[3]{Ulrich Heber}

  \affil[1]{Dr. Karl Remeis-Observatory \& ECAP, Astronomical Institute, Friedrich-Alexander University Erlangen-Nuremberg, Sternwartstr. 7, 96049 Bamberg, Germany, E-mail: matti.dorsch@fau.de}
  \affil[2]{Dr. Karl Remeis-Observatory \& ECAP, Astronomical Institute, Friedrich-Alexander University Erlangen-Nuremberg, Sternwartstr. 7, 96049 Bamberg, Germany, E-mail: marilyn.latour@fau.de}
  \affil[3]{Dr. Karl Remeis-Observatory \& ECAP, Astronomical Institute, Friedrich-Alexander University Erlangen-Nuremberg, Sternwartstr. 7, 96049 Bamberg, Germany, E-mail: ulrich.heber@sternwarte.uni-erlangen.de}

  \title{\huge Spectral analysis of the He-enriched sdO-star HD\,127493}

  \runningtitle{Spectral analysis of HD\,127493}
  \begin{abstract}
{
The bright sdO star HD\,127493 is known to be of mixed H/He composition and excellent archival spectra covering both optical and ultraviolet ranges are available. UV spectra play a key role as they give access to many chemical species that do not show spectral lines in the optical, such as iron and nickel. This encouraged the quantitative spectral analysis of this prototypical mixed H/He composition sdO star.
We determined atmospheric parameters for HD\,127493 in addition to the abundance of C, N, O, Si, S, Fe, and Ni in the atmosphere using non-LTE model atmospheres calculated with \texttt{TLUSTY}/\texttt{SYNSPEC}. A comparison between the parallax distance measured by Hipparcos and the derived spectroscopic distance indicate that the derived atmospheric parameters are realistic. From our metal abundance analysis, we find a strong CNO signature and enrichment in iron and nickel.
}
\end{abstract}

  \journalname{Open Astronomy}
\DOI{DOI}
  \startpage{1}
  \received{Oct 9, 2017}
  \revised{..}
  \accepted{Nov 21, 2017}

  \journalyear{2017}
  \journalvolume{1}
\maketitle
\section{Introduction}
\vspace*{-5pt}
Hot subluminous stars come in two flavors: sdO and sdB stars. All of them display peculiar helium abundances; they are either deficient or enriched in helium. The vast majority of sdB stars are helium deficient, while the helium enrichment of the small group of He-sdBs is mostly moderate, their atmospheres are of mixed H/He composition. A different pattern emerges for sdO stars. Most of them have atmospheres dominated by helium with hydrogen being a trace element only and, therefore, are termed He-sdO. Helium-poor sdOs exist too, though at lower numbers. Mixed composition objects, however, are rare among sdO stars {\color{Black}\citep{Nemeth12}}. HD\,127493 is a bright He-sdO star and was among the first hot subdwarfs to be discovered \citep{munch55}. Its parallax was measured as $5.43\pm 1.21$\,mas by the Hipparcos satellite \citep{vanleeuwen07}, which corresponds to a distance of $184^{+53}_{-34}$\,pc. 
Because of its brightness ($m_V=10.08$), it has been analyzed a few times in past studies.
\cite{simon82} estimated an effective surface temperature $T_\mathrm{eff}$ of $42\,500$\,K, a surface gravity of $\log g = 5.25$ and a moderately enriched helium abundance ($\log N(\mathrm{He})/N(\mathrm{H})=0.60$). \cite{hirsch09} reanalyzed the star using new spectra and more sophisticated model atmospheres calculated with the \texttt{TMAP} model code \citep{werner99,werner03}. He found the star's atmosphere to be enhanced in nitrogen, while carbon is depleted. 
This implies that large amounts of nitrogen from the CNO cycle have reached the surface \citep{bauer95}.
Most hot subdwarf stars are either He-rich or He-weak, while few show a N(He)/N(H) ratio close to one \citep[Fig.~4]{fontaine14,heber16}. 
Given its brightness and CNO abundances, this makes HD\,127493 an interesting target for a more detailed analysis. 
%
%
The aim of this work is to derive the chemical abundance of all relevant metals (like carbon, nitrogen, oxygen, silicon, sulfur, iron, and nickel) in the star's photosphere, as well as to revisit the atmospheric parameters determined by \cite{hirsch09} and \cite{simon82} with more sophisticated model atmospheres. 
To this end, we analyzed high quality UV spectra of HD\,127493 taken with the HST GHRS instrument for the first time, as well as high resolution FEROS spectra. 
A similar analysis has been performed before for another mixed He-composition sdO star, BD\,+75\,325, by \cite{lanz97}. 
\smallskip\\
Such an abundance analysis can give important clues regarding the evolutionary history of the star. For single He-sdOs, there are two main evolutionary scenarios: a merger of two helium white dwarfs \citep[He-WDs,][]{webbink84,zhang12a} or a late He-flash, where the subdwarf progenitor undergoes significant mass loss at the top of the red giant branch and ignites its He-core while contracting on the white dwarf cooling sequence \citep{castellani93,miller08}. These scenarios can both produce He-sdOs, albeit with different surface metal abundances.
\vspace*{-35pt}
\section{Archival spectra}
\vspace*{-5pt}
\begin{table}[!htbp]
\caption{List of observations used in our analysis.}
\vspace*{-10pt}
\label{tab:spectra}
\begin{center}
\resizebox{1\columnwidth}{!}{%
\setlength\extrarowheight{2pt}
\begin{tabular}{c c c c c}\hline
Instrument	&	Dataset	&	 Range (\AA) 	&Exp. (s)	\\ \hline\hline 
FEROS		&	ADP.2016-09-21T07:07:18.680		&	$3527.9-9217.7$	&	$\phantom{0}600$	\\
			&	ADP.2016-09-21T07:07:18.736		&	$3527.9-9217.7$	&	$\phantom{0}300$	\\
			&	ADP.2016-09-21T07:07:18.686		&	$3527.9-9217.7$	&	$\phantom{0}300$	\\ \hline
HST GHRS	&	Z2H60107T	&	$1222.6-1258.8$	&	$\phantom{0}462$		\\	
			&	Z2H60109T	&	$1254.9-1291.0$	&	$\phantom{0}462$		\\
			&	Z2H6010BT	&	$1285.6-1321.6$	&	$\phantom{0}462	$	\\
			&	Z2H6010DT	&	$1317.7-1353.6$	&	$\phantom{0}462	$	\\
			&	Z2H6010FT	&	$1349.7-1385.5$	&	$\phantom{0}516	$	\\	
			&	Z2H6010HT	&	$1383.0-1418.8$	&	$\phantom{0}598	$	\\	
			&	Z2H6010JT	&	$1414.9-1450.5$	&	$\phantom{0}516	$	\\	
			&	Z2H6010LT	&	$1532.5-1567.7$	&	$\phantom{0}652	$	\\	
			&	Z2H6010OT	&	$1623.2-1658.1$	&	$\phantom{0}462	$	\\	
			&	Z2H6010QT	&	$1713.0-1747.6$	&	$\phantom{0}462	$	\\ \hline 
IUE			&	SWP04860	&	$1150.0-1980.0$	&	$3000	$	\\
			&	SWP07695	&	$1150.0-1980.0$	&	$4500	$	\\
			&	SWP08276	&	$1150.0-1980.0$	&	$3930$		\\ \hline
\end{tabular}

}
\end{center}
\vspace*{-20pt}
\end{table}
\noindent We used optical FEROS spectra to re-determine the atmospheric parameters of HD\,127493. FEROS is an echelle spectrograph mounted at the MPG/ESO-2.20m telescope operated by the European Southern Observatory (ESO) in La Silla. It features a high resolving power of $\mathrm{R}\approx 48000$ \citep{kaufer99} and its spectral range from $\sim$\,3600\,\AA\ to $\sim$\,9200\,\AA\ includes all Balmer lines as well as many He\,{\sc i}, He\,{\sc ii}, nitrogen, and silicon lines.
The three available spectra of HD\,127493 were co-added to achieve a high signal-to-noise ratio (S/N) of $\gtrsim$\,100 in the $4000-6000$\,\AA\ range. Towards both ends of the spectral range, the S/N decreases drastically, reaching a value of S/N\,$\sim$\,15 at 8000\,\AA .
\smallskip \\
Observations in the UV spectral range are very useful for HD\,127493, since this is the range where the star's flux is highest. 
In addition, strong metal resonance lines and a large number of iron and nickel lines make UV spectra a valuable tool for measuring the chemical surface composition of the star.
Interstellar lines at short wavelengths are rather well known since they are frequently seen in UV spectra {\color{Black}and usually have} low ionization stages, so they could be easily identified.
\smallskip \\
We retrieved three archival spectra from the International Ultraviolet Explorer (IUE) from the MAST archive\footnotemark and co-added them to achieve a good S/N ratio.
They continuously cover the $1150-1980$\,\AA\ range with a resolution of about 0.15\,\AA . This range includes  C\,{\sc iii} resonance and N\,{\sc iii} lines that are important for the abundance analysis.
\smallskip\\
HD\,127493 has also been observed with the Goddard High-Resolution Spectrograph (GHRS) mounted on the Hubble Space Telescope (HST). These publicly available spectra cover the $1225-1745$\,\AA\ range with a resolution of about 0.06\,\AA\ (medium resolution mode, R $\approx 25000$). The S/N is very high, especially around the middle of the covered range. The quality of this spectrum is better than that from IUE, both with respect to S/N and spectral resolution.
\vspace*{-10pt}
\section{Atmospheric parameters}
In order to analyze these spectra, we computed non-LTE model atmospheres using the \texttt{TLUSTY} and \texttt{SYNSPEC} codes developed by \cite{hubeny88}. A detailed description of \texttt{TLUSTY} is given in the form of a three-part manual, see \cite{hubeny17a,hubeny17b,hubeny17c}.
\smallskip \\
Previous spectral analyses of HD\,127493 were based on different and less sophisticated model atmospheres.
For that reason we found it necessary to re-determine the atmospheric parameters before embarking on the metal abundance determination.
\footnotetext{Mikulski Archive for Space Telescopes, \url{https://archive.stsci.edu/index.html}}
We computed a small 3D grid for fitting the optical spectrum, including effective temperatures of $T_\mathrm{eff}=35\,000$, 40\,000, and 45\,000\,K, as well as surface gravities of $\log g =$ 4.7, 5.4, and 6.1.
For each of these combinations, models with a helium abundance of $\log N(\mathrm{He})/N(\mathrm{H})=-0.3$, +0.3, and +0.9 were computed. 
In addition to hydrogen and helium, all models in the grid include carbon, silicon and nitrogen in non-LTE with the abundance as determined by our analysis.
The selection of all lines we used, as well as the global best-fit model can be seen in Fig. \ref{fig:atm_fit}. The results from a simultaneous fit of all selected lines are shown in Table \ref{table:tab_abu}
\footnote{We are aware that this grid is coarse and it can not be excluded that interpolation is not accurate enough. The grid would need to be extended to rule out interpolation issues. Hence, these results are preliminary and we have not carried an investigation of the uncertainties. Nevertheless, they agree well with the results of \cite{hirsch09} obtained with \texttt{TMAP}.}.
Therefore, we adopted the atmospheric parameters of \cite{hirsch09} for our fully line-blanketed models.
\begin{minipage}{1\columnwidth}
\vspace*{12pt}
\captionof{table}{Results from the new 3D-fit compared to the results of \cite{hirsch09} with error estimates from spectrum-to-spectrum variations.}\label{table:tab_abu}
\vspace*{-15pt}
\begin{center}
\resizebox{\columnwidth}{!}{%
\begin{tabular}{l||c|c}
					&H/He/C/N/Si & H/He/N 		\\
					& (this work) 	& \citep{hirsch09}\\ \hline 
$T_\mathrm{eff}\,(\mathrm{K})$ 	&$41\,900 $ & $42\,484 \pm 250$\\
$\log g$			&$5.561$& $5.60\pm 0.05$	\\
$\log N(\mathrm{He})/N(\mathrm{H})$			&$0.435$	& $0.62\pm 0.30$	\\
\end{tabular}

}
\end{center}
\vspace*{-10pt}
\end{minipage}
\begin{minipage}{1\columnwidth}
\centering
    \includegraphics[width=0.9\columnwidth]{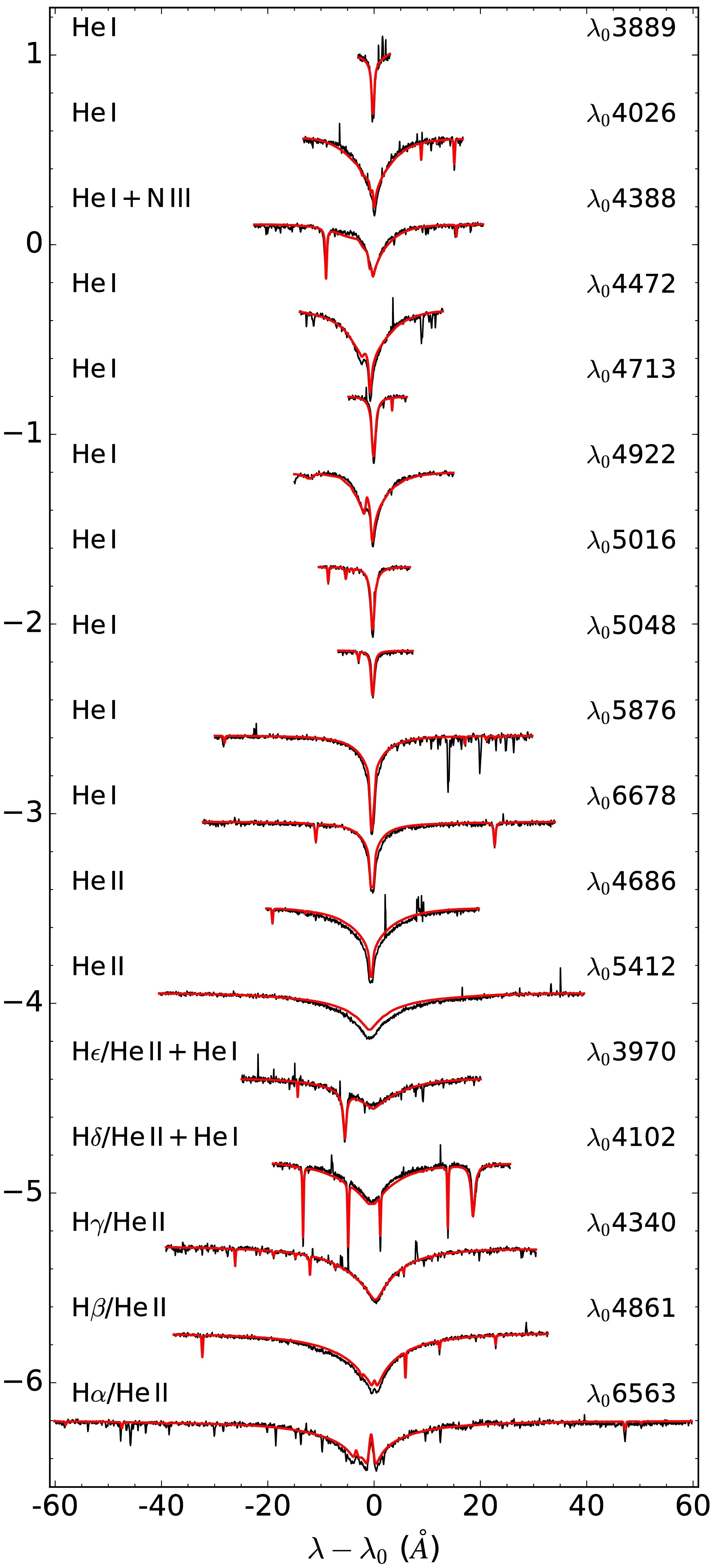}
    \captionof{figure}{Best fit (red) to the optical Balmer and helium lines selected in the normalized FEROS spectra (black).}\label{fig:atm_fit}
\end{minipage}\\
{\color{Black}\section{Spectroscopic distance}}
Since a Hipparcos parallax (and therefore distance) measurement is available for HD\,127493, we checked for consistency by comparing it to the spectroscopic distance derived from our model atmosphere when assuming a canonical stellar mass ($0.48\,M_\odot$). 
The absolute $V$-magnitude ($M_V$) derived from a model atmosphere was combined with the apparent $V$-magnitude ($m_V$) of the star to derive its distance. First, the apparent magnitude was corrected for interstellar reddening, so that it can be compared to the theoretical unreddened magnitude. Interstellar dust, containing large amounts of metals, scatters and absorbs light especially in the ultraviolet through infrared and emits photons thermally in the far-infrared \citep{fitzpatrick99}. Since the dust extinction is stronger at shorter wavelengths, this leads to a reddening effect, which can be described by the difference between the observed and intrinsic color index of a star (usually the $B-V$ color): 
$E_{B-V}=(B-V)_\mathrm{observed}-(B-V)_\mathrm{intrinsic}$.
The overall reddening along the line-of-sight in the direction of HD\,127493 is listed in the Infrared Science Archive\footnote{\url{https://irsa.ipac.caltech.edu/applications/DUST/}} using data from \cite{schlafly11} as $E_{B-V}=0.0801 \pm  0.0006$. This can be seen as an upper limit on the reddening. 
The theoretical $V$\,$-$\,flux (at 5454\,\AA ) was obtained from interpolation between a grid of synthetic non-LTE spectra to match the effective temperature and surface gravity determined from our analysis. To compute the absolute magnitude of the star from the theoretical model flux, the flux was scaled with the star's radius, which can be calculated from the surface gravity $g$ and mass $M$: $R=\sqrt{GM/g}$, where $G$ is the gravitational constant.
After correcting the apparent magnitude for reddening, the distance was calculated by simply using the distance modulus: $d_\mathrm{spectroscopic}=10^{1+\,(m^0_V-M_V)/5 }$.
When the total line-of-sight reddening is assumed, the resulting spectroscopic distance is $d_\mathrm{spectroscopic}=178\,\pm\, 32$\,pc, which provides a lower limit for the spectroscopic distance. 
This value agrees reasonably well with the parallax distance $d_\mathrm{parallax}=184^{+53}_{-34}$\,pc, indicating that both the assumption of mass as well as the determined surface gravity are appropriate.
\begin{table*}[!htbp]
\centering
\caption{{Abundance results by number ($\log N_X/N_\mathrm{H}$) and mass fraction ($\beta _X$) with standard deviation errors from single lines.}}\label{tab_metal}
\resizebox{2\columnwidth}{!}{%
{\setlength{\extrarowheight}{2pt}%
\begin{tabular}{l||c|c|c|c|c|c|c|c|c}
			& 	H	&	He	&	C	&	N	&	O	&	Si	&	S	&	Fe	&	Ni	\\  \hline  
$\log N_X/N_\mathrm{H}$			& 	$0$	&	$+0.44^{+0.10}_{-0.12}$	&	$-4.62^{+0.07}_{-0.08}$	&	$-2.38^{+0.13}_{-0.18}$	&	$\leq -4.62$	&	$-3.63^{+0.12}_{-0.16}$	&	$-4.01^{+0.19}_{-0.34}$	&	$-3.20^{+0.14}_{-0.21}$	&	$-3.97^{+0.06}_{-0.07}$	\\
$\log \beta _X$			& 	$-1.08^{+0.09}_{-0.11}$	&	$-0.04^{+0.13}_{-0.18}$	&	$-4.62^{+0.11}_{-0.15}$	&	$-2.32^{+0.15}_{-0.23}$	&	$\leq -4.50$	&	$-3.26^{+0.14}_{-0.21}$	&	$-3.59^{+0.20}_{-0.39}$	&	$-2.54^{+0.16}_{-0.25}$	&	$-3.29^{+0.11}_{-0.14}$	\\
\end{tabular}
}
}
\end{table*}
\vspace*{-3pt}
\begin{figure*}[htbp]
\begin{minipage}{0.7\textwidth}
\begin{center}
	\includegraphics[width=1\textwidth]{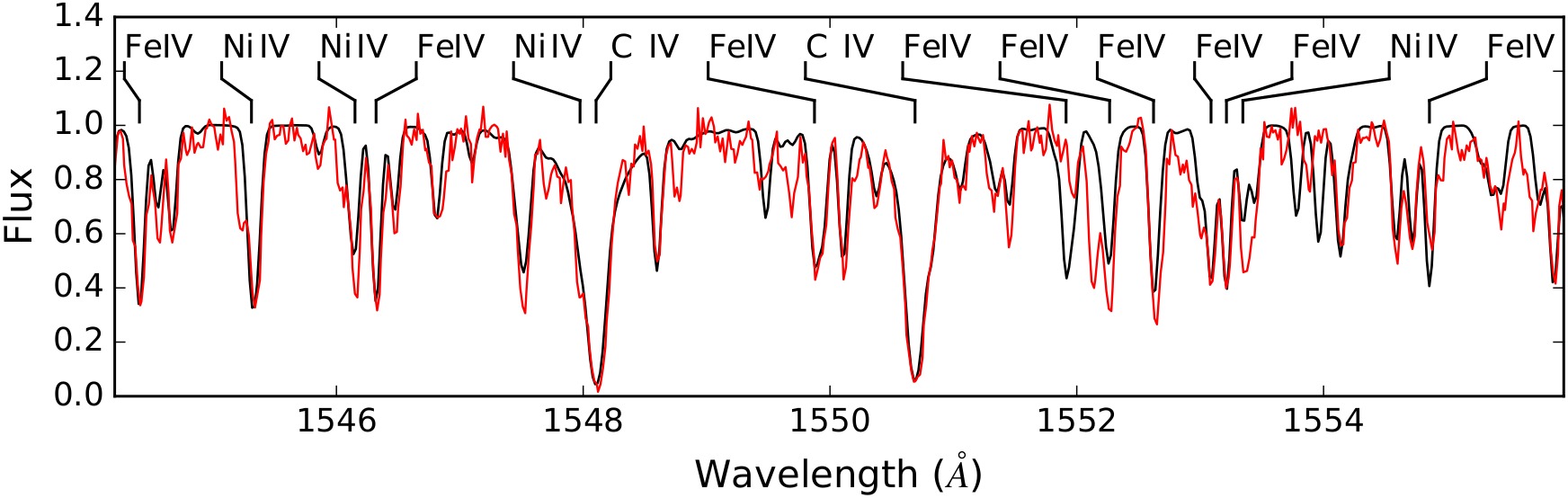}\\
	\includegraphics[width=1\textwidth]{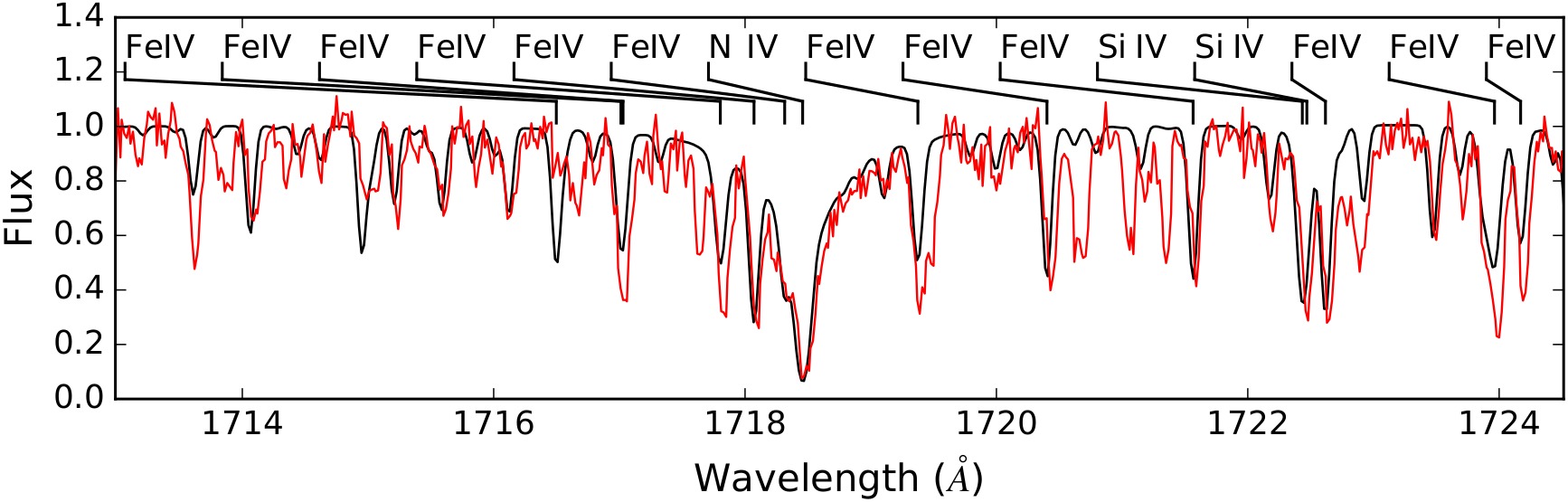}
	\captionof{figure}{Two example spectral ranges in the GHRS spectra (red), compared to the best-fit model spectrum in black.}
	\vspace*{-13pt}
	\label{nUV}
\end{center}
\end{minipage}
\hspace*{4pt}
\begin{minipage}{0.3\textwidth}
    \includegraphics[width=1\columnwidth]{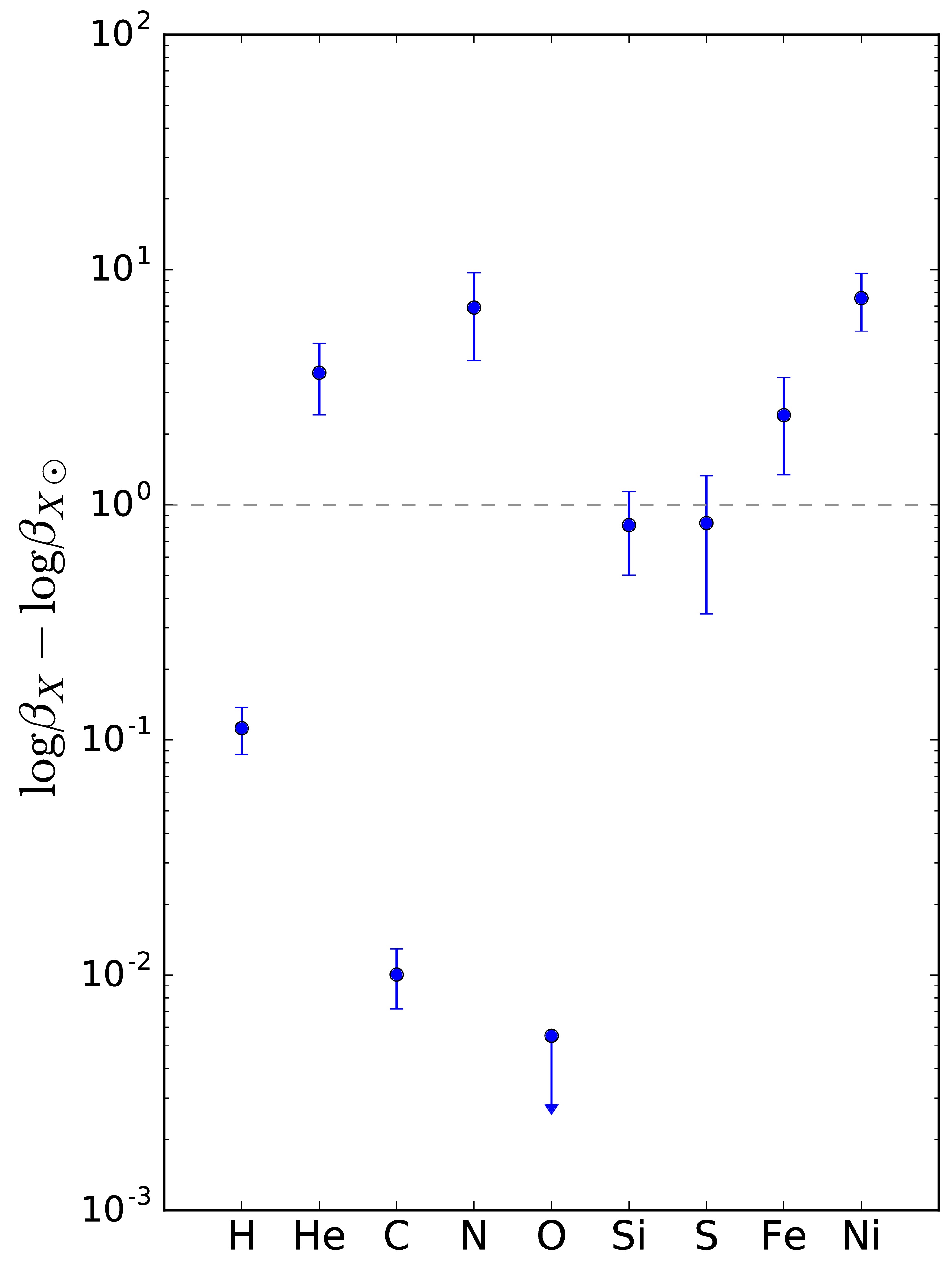}
    \captionof{figure}{Chemical surface composition in solar units. Values for the solar abundances were adopted from \cite{asplund09} and \cite{scott15a,scott15b} for elements heavier than Ca.}\label{fig:solar}
\end{minipage}
\end{figure*}

\section{Abundance analysis}
In order to derive the chemical abundances we used fully line-blanketed non-LTE models. The final model includes the following ions in non-LTE: H, He\,{\sc i}-{\sc ii}, C\,{\sc iii}-{\sc iv}, N\,{\sc ii}-{\color{Black}{\sc v}}, O\,{\sc ii}-{\sc v}, Si\,{\sc ii}-{\sc iv}, S\,{\sc iii}-{\sc v}, Fe\,{\sc iii}-{\sc v}, Ni\,{\sc iv}-{\sc v}, for a total of 877 energy levels overall. 
{\color{Black} The highest ionization stage of every metal is considered as a one-level ion.} 
For all other ionization stages we used the most detailed model atoms available on the \texttt{TLUSTY} website\footnote{\url{http://nova.astro.umd.edu/Tlusty2002/tlusty-frames-data.html}}. Starting from a pure H/He model, additional elements were added to the model one at a time to ensure convergence. 
\vspace*{-10pt}
\newpage
\noindent We determined the abundance of metals in the atmosphere of HD\,127493 using one-dimensional grids with fixed $T_\mathrm{eff}$, $\log g$, and helium abundance for each chemical element by fitting pre-selected spectral lines individually.
%
%
We found most line profiles to be well reproduced with a projected rotational velocity below 10\,km\,s$^{-1}$.
Hence, HD\,127493 is a slow rotator unless its inclination is low.
\smallskip \\
The optical FEROS spectrum shows N\,{\sc ii}-{\sc iv} and Si {\sc iii}-{\sc iv} lines, most of which could be used for fitting. For lines of other elements, the UV spectra were used. 
Here, we started with determining the abundance of iron and nickel, since lines from these two elements dominate this spectral range (see Fig. \ref{nUV}). 
Despite the large number of blended lines, UV spectra are a valuable tool for determining the abundance of light metals, as carbon, nitrogen and silicon have strong lines in this range. The C\,{\sc iv} resonance doublet shown in the top panel of Fig.  \ref{nUV} and the strong N\,{\sc iv} line in the bottom panel are two examples.
{\color{Black}The abundances derived using different ionization stages of a given atom (C\,{\sc iii}-{\sc iv}, N\,{\sc iii}-{\sc v}, Si\,{\sc iii}-{\sc iv}, S\,{\sc iv}-{\sc v}, Fe\,{\sc iv}-{\sc v}, and Ni\,{\sc iv}-{\sc v}) are consistent, indicating that the temperature used in our models is accurate.}
\smallskip \\
Our resulting chemical composition is summarized in Table \ref{tab_metal}. Note that the abundance \textit{by number} is relative to the hydrogen abundance, which is below solar. For better comparison to other He-enriched stars, the abundance of every chemical element $X$ is also listed in mass fraction $\beta _X$. Uncertainties are given as standard deviation from single-line fits. For most elements, multiple ionizion stages could be used to check the effective temperature considering the ionization balance, which matches very well.
\smallskip\\
From a comparison to solar values (Fig.~\ref{fig:solar}), HD\,127493 can be categorized as an intermediate He-sdO of the N-type. Carbon and oxygen are depleted, as expected from CNO-processed material. Silicon and sulfur abundances are close to solar.  Iron is enriched to about 1.5 to 3.5 times solar, while nickel is enriched to about 5 to 10 times solar. 
\vspace*{-5pt}
\section{Model atmosphere characteristics}
When including additional chemical elements, mostly the outer layers of the atmosphere are affected by the increased opacity created by numerous metal absorption lines in the UV range (see Fig.~\ref{depth}, top panel). This additional opacity blocks outgoing flux, so the continuum flux has to rise in order to ensure flux conservation. This requires a steeper temperature gradient in the region where the continuum originates, leading to higher temperatures (the so-called back-warming effect). At the same time, the temperature is lowered in the outer atmosphere, where the cores of the strong metal lines become optically thin (surface cooling).
Both effects become apparent in Fig.~\ref{depth}b when comparing the temperature structure of the model with nitrogen only (blue) to the fully line-blanketed one (red): adding more elements to the model lowers the temperature at column masses below $\sim\,10^{-2}$\,g\,cm$^{-2}$, whereas deeper regions are heated. Even adding only nitrogen to a H/He only model results in a large temperature difference for column densities less than $10^{-3}$\,g\,cm$^{-2}$. 
\smallskip \\
Fig.~\ref{depth} shows at which column densities in the atmosphere the monochromatic optical depth $\tau _\nu = 2/3$ is reached (i.\,e. where about half of the photons escape the photosphere). In combination with the temperature stratification profile, this makes it possible to analyze how single lines are affected by the temperature at a certain column density. The line cores of many metal lines in the UV range form close to the surface, while many optical lines, most importantly many helium and hydrogen lines, are created deeper in the atmosphere, between $10^{-1}$ and $10^{-2}$\,g\,cm$^{-2}$. This region is strongly affected by the back-warming effect.
\smallskip \\
Since the metal ionization fractions as a function of depth strongly depend on $T_\mathrm{eff}$, observing more than one ionization stage for some elements is useful to constrain this parameter. 
How sensitive metal ionization equilibria are to the temperature is shown in the bottom panel of Fig.~\ref{ion_1} using the example of N\,{\sc ii}-{\sc vi}. While the line-forming region around $10^{-2}$\,g\,cm$^{-2}$ is dominated by N\,{\sc iii} in the $40\,000\,$K model, the N\,{\sc iii}/N\,{\sc iv} ratio is almost inversed in the $45\,000\,$K model. 
\smallskip \\
The top panel of Fig.~\ref{ion_1} displays the effect of line-blanketing on the ionization structure of nitrogen by comparing a H/He/N model atmosphere to the fully line-blanketed one. 
As expected from the change in temperature stratification, the population of N\,{\sc iv}-{\sc vi} is increased in the line-blanketed model for high column densities, while it is lowered towards the surface. The reverse effect is seen for N\,{\sc iii}. Since the differences can be quite large, especially in line-forming regions, this underlines the importance of using line-blanketed models deriving abundances (even in the optical range).
\begin{minipage}{1\columnwidth}
\centering
    \includegraphics[width=1\columnwidth]{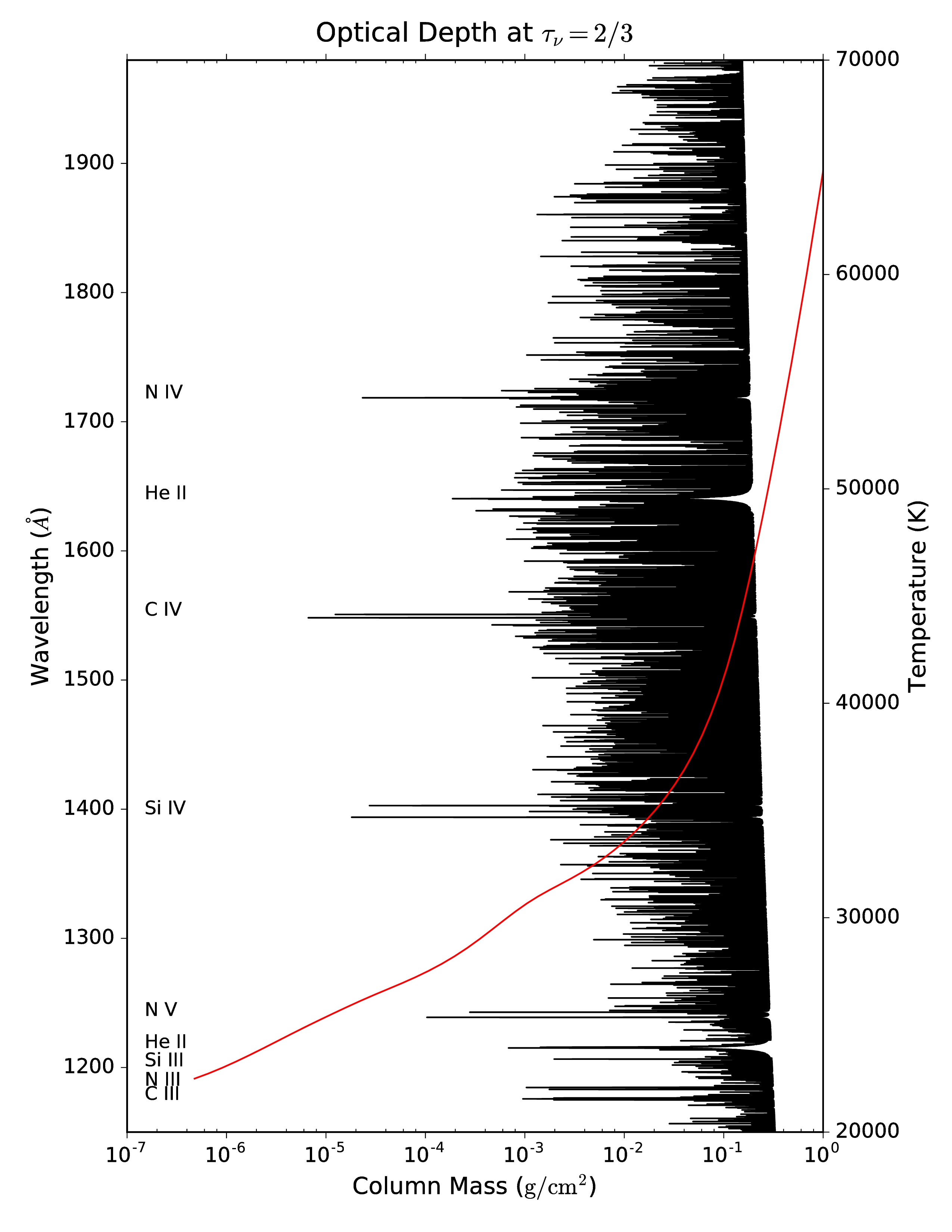}\\
    \includegraphics[width=1\columnwidth]{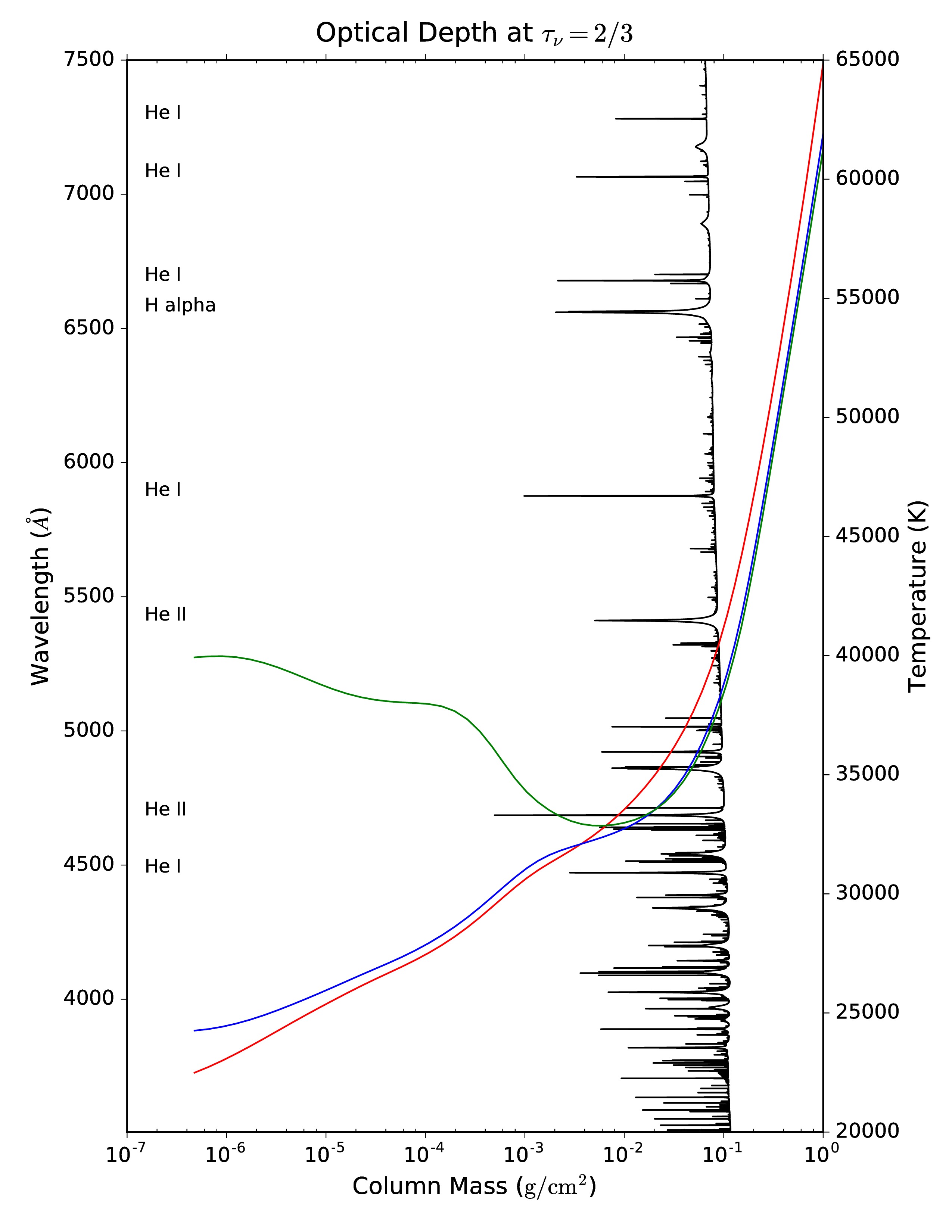}
		\captionof{figure}{Top: temperature stratification and monochromatic optical depth $\tau _\nu = 2/3$ as functions of column mass density for the final fully line-blanketed  model defined by $T_\mathrm{eff} = 42\,484$\,K, $\log g = 5.6$, and $\log N(He)/N(H) = 0.62$. Bottom: The temperature stratification for models including H+He only (green), N (blue) and a fully line-blanketed model (red) is shown in the optical range, as well as $\tau _\nu = 2/3$.}
    \label{depth}
\end{minipage}
\begin{minipage}{1\columnwidth}
    \includegraphics[width=1\textwidth]{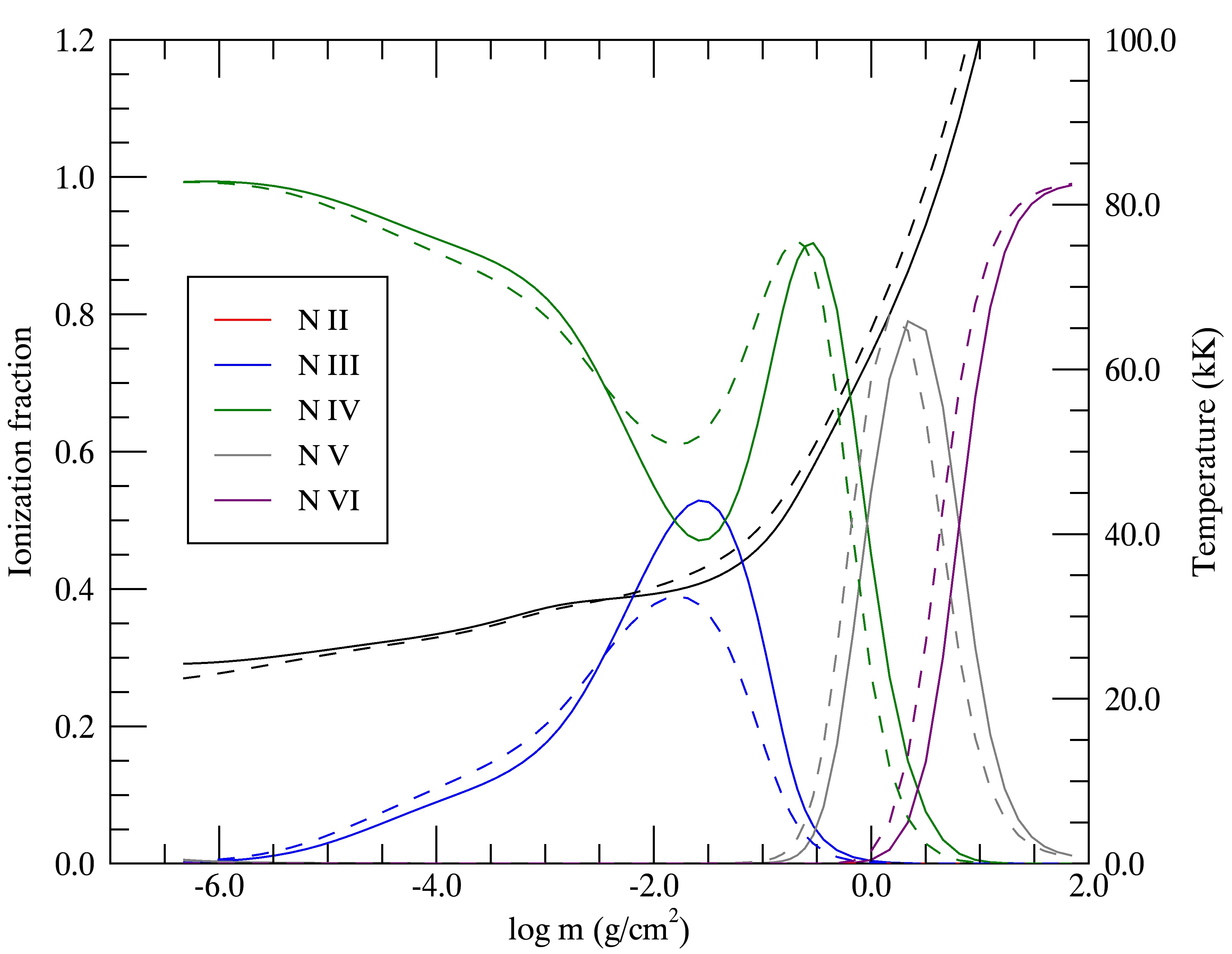}
    \includegraphics[width=1\textwidth]{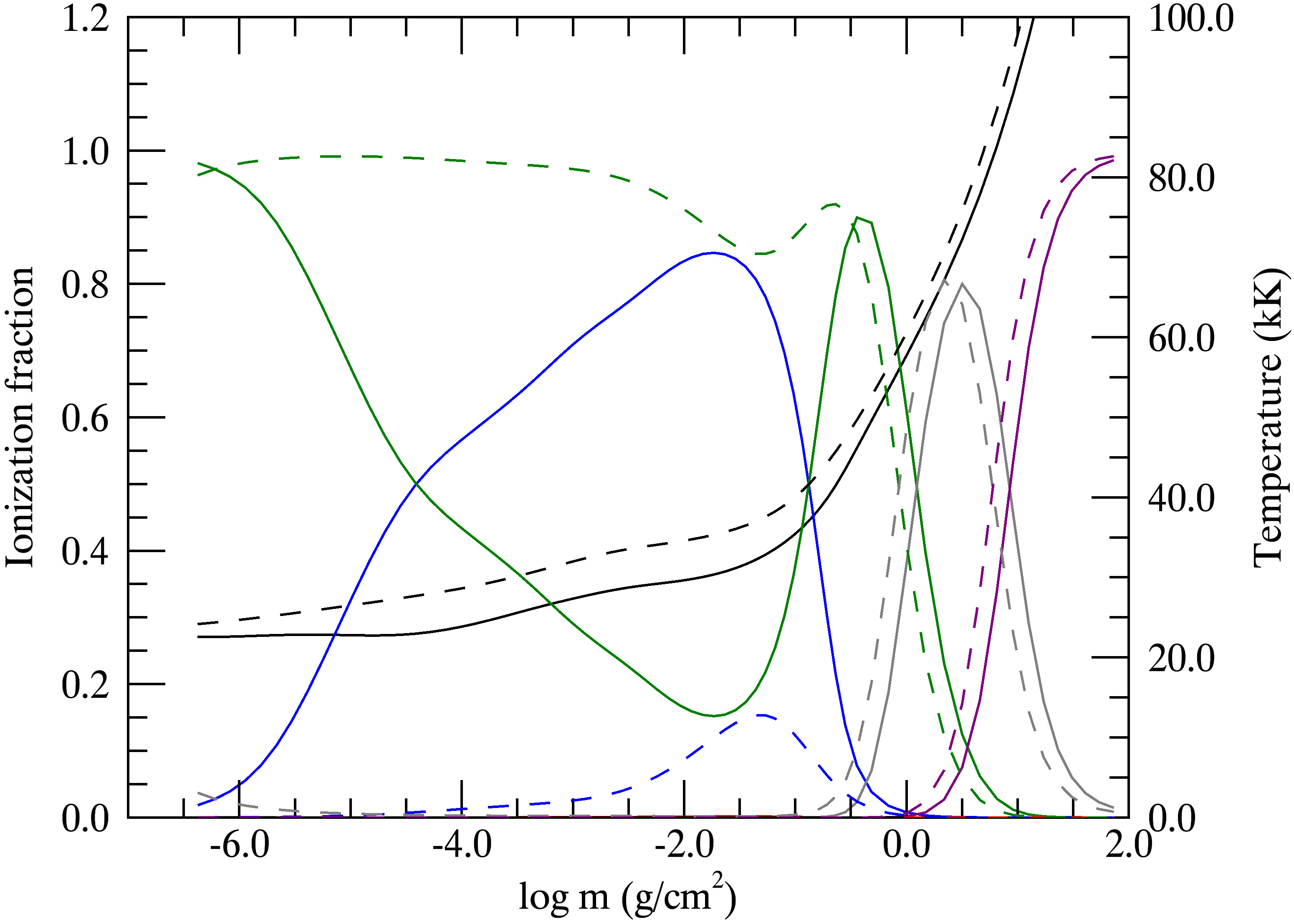}
    \captionof{figure}{Top: Nitrogen ionization fractions  in H/He/N models (solid) and models including C/O/Si/S/Fe/Ni also (dashed lines) and the temperature stratification (black). Bottom: Nitrogen ionization fractions in models with $T_\mathrm{eff}=40\,000$\,K (solid) and $45\,000$\,K (dashed) and the temperature stratification (black).}\label{ion_1}
\end{minipage}
\vspace*{-0pt}
\section{Discussion and outlook}
The atmospheric parameters for HD\,127493 were revisited using a small new grid of models and synthetic spectra computed with \texttt{TLUSTY}/\texttt{SYNSPEC}.
Our results confirm previous literature values and the spectroscopic distance matches the Hipparcos parallax distance very well, further validating the derived surface gravity and assumed canonical mass.
The HST-GHRS UV-spectra turned out to be especially useful for determining metal abundances and could be reproduced well with the final synthetic spectrum. 
The atmosphere of HD\,127493 is enriched in nitrogen but depleted in carbon, whereas the abundances of silicon and sulfur are almost solar. Interestingly, both iron and, most notably, nickel are enriched compared to solar values. 
\smallskip \\
{\color{Black} 
In addition, we found no evidence for a companion. All analyzed spectra  (taken in 1979-1980, 1996, and 2005) are consistent with a radial velocity of about $-18\pm 5$\,km\,s$^{-1}$. 
We compared the spectral energy distribution of our best-fit model with the photometric measurements available and did not detect any sign of an infrared excess.
Nevertheless, a low-mass companion can not be ruled out since no extensive radial velocity study has yet been performed for HD\,127493.}
\smallskip \\
Based on these results, a hot flasher origin for HD\,127493 can be considered unlikely. 
%
%
Models by \cite{miller08} predict He-sdOs enriched in carbon, as opposed to what is observed for HD\,127493. 
%
On the other hand, the slow and low-mass composite merger scenarios of two He-WD presented by \cite{zhang12a} are able to produce He-rich sdOs with a high abundance of nitrogen{\color{Black}, but they also predict a higher oxygen abundance than what we measured in the star}.
Also the rotational velocity of a He-WD+He-WD merger product would be orders of magnitude higher than the observed $v_\mathrm{rot} \sin i$ below 10\,km\,s$^{-1}$, if no angular momentum were lost \citep{gourgouliatos06}.
In addition, semi-convective or rotational mixing could strongly influence the chemical abundances in the atmosphere. The same holds for gravitational settling and radiative levitation, which might be able to explain the enrichment in iron and nickel in HD\,127493.
\smallskip \\
In the future, the determination of atmospheric parameters of HD\,127493 could be further improved by using a fine grid of line-blanketed atmospheres. In addition, GAIA satellite parallax measurements will be a crucial test for spectroscopic distance measurements and, therefore, stellar mess. 
Finally, detailed evolutionary scenarios, including a more complete treatment of the atmosphere, would provide better predictions for observed surface abundances.
\vspace*{-1pt}
\section*{Acknowledgements}
Based on observations made with the NASA/ESA Hubble Space Telescope, obtained from the data archive (prop. ID GO5305) at the Space Telescope Science Institute. STScI is operated by the Association of Universities for Research in Astronomy, Inc. under NASA contract NAS 5-26555. Support for MAST for non-HST data is provided by the NASA Office of Space Science via grant NNX09AF08G and by other grants and contracts. 
Based on observations made with ESO Telescopes at the La Silla Paranal Observatory under programme ID  074.B-0455(A).
\smallskip\\
M.\,L. acknowledges support from the Alexander von Humboldt Foundation.
\bibliographystyle{apj}
\bibliography{bsc_proc_old}
%
%
%
\end{document}